\documentclass[manuscript]{aastex}
\usepackage{graphicx}
\usepackage{color}
\begin{document}

\shorttitle{Multi-wavelength Emission .....}
 \shortauthors{Cheng et al.}

\title{Multi-wavelength Emission from the Fermi Bubble III. Stochastic (Fermi) Re-Acceleration of Relativistic Electrons Emitted by SNRs.}
\author{K. S. Cheng$^{1}$, D. O. Chernyshov$^{1,2}$, V. A. Dogiel$^{1,2,3}$,  and C. M. Ko$^{4}$}
\affil{$^1$Department of Physics,
University of Hong Kong, Pokfulam Road, Hong Kong, China}
\affil{$^2$I.E.Tamm Theoretical Physics Division of P.N.Lebedev
Institute of Physics, Leninskii pr. 53, 119991 Moscow, Russia}
\affil{$^3$Moscow Institute of Physics and Technology (State University), 9, Institutsky lane, Dolgoprudny, 141707, Russia}
\affil{$^4$Institute of Astronomy, Department of Physics and Center for Complex Systems, National Central University,
Jhongli, Taiwan}



\altaffiltext{0}{........}

\begin{abstract}
We analyse the model of stochastic re-acceleration of electrons, which are  emitted by supernova remnants (SNRs)
in the Galactic Disk and propagate then into the Galactic halo, in order to explain the origin on nonthermal
(radio and gamma-ray) emission from the Fermi Bubbles (FB). We assume that the energy for re-acceleration in the halo
is supplied by shocks generated by processes of star accretion  onto the central black hole.  Numerical simulations
show that regions with strong turbulence (places for electron re-acceleration) are located high up in the Galactic
Halo about several kpc above the disk. The energy of SNR electrons that reach these regions does not exceed several
GeV because of synchrotron and inverse Compton energy losses. At appropriate parameters of re-acceleration these
electrons can be re-accelerated up to the energy $10^{12}$ eV which  explains in this model the origin of the observed
radio and gamma-ray emission from the FB.
However although the model gamma-ray spectrum is consistent with the Fermi results, the model radio spectrum is steeper
than the observed by WMAP and Planck. If  adiabatic losses due to plasma outflow from the Galactic central regions are
taken into account, then the re-acceleration model nicely reproduces the Planck datapoints.
\end{abstract}


\date{\today}

\maketitle

\section{Introduction}
Recently Fermi has discovered two giant gamma-ray Bubbles (FBs)
that extend nearly 10 kpc in diameter north and south of the
Galactic center \citep[cf.][for more recent analyses, see \citet{hoop,Yang2014,acker14}]{dob10,meng}. These
gamma-ray Bubbles also correlate with the earlier discovered
so-called "microwave haze" observed by the WMAP telescope as
described by \citet{fink} and \citet{dob08},
and with the large scale X-ray emission region first evidenced by
analysing the ROSAT 1.5 keV data, which clearly showed the
characteristic of a bipolar flow \citep[see e.g.,][]{Snowden,cohen}. A number of models
was suggested to explain the origin of the FBs either  due to protons-proton collisions
\citep[hadronic model, see e.g.,][]{crock11,crocker14a,crocker14b}
or due to inverse Compton scattering of relativistic electrons
\citep[leptonic model, see e.g.,][]{meng}.

Several requirements follow for the leptonic models from observations. First, the gamma-ray emission has a cut-off at
$E_\gamma\sim 100$ GeV \citep[see][]{meng, Yang2014}. Then, in the case of inverse Compton origin of the FB gamma-rays
electrons have to be accelerated there up to the energy
\begin{equation}
E^e_{max}\la m_ec^2\sqrt{\frac{3}{4}\frac{E_\gamma}{\varepsilon}}\sim 5\times 10^{11}~\mbox{eV}\,,
\end{equation}
where $\varepsilon\simeq 10^{-3}$eV is the energy of the microwave photons.

Secondly,  because of a very short lifetime of electrons, they have to be in-situ generated in regions of emission.
\citet{cheng} assumed that electrons are accelerated at the FB edge by shocks while in the model of \citet{mertsch}
it was assumed that  this emission is generated by electrons  accelerated in the Galactic halo
by an MHD-turbulence which is excited by a shock propagating into the halo.
The energy of this shock cascades into turbulence by different processes of plasma instabilities.
Interaction of electrons with this turbulence leads to stochastic or second-order Fermi acceleration.
Alternatively particles can  be stochastically accelerated in the FBs by interaction with a
supersonic turbulence (shocks) which is excited by tidal processes in the Galactic Center \citep[][]{cheng12}.

These models were investigated in the test particle approximation when feedback reaction of accelerated particles on
the acceleration mechanism is ignored. Therefore, the number of accelerated particles is usually a free parameter of the models.
However, in some cases when the sources of accelerated electrons are known this number can be estimated from kinetic equations
that gives additional model restrictions. In the case of FBs there are three evident sources of electrons. The electrons
can be supplied by: a) Coulomb collisions from the FB background plasma, b) by $pp$ collisions in the halo (secondary electrons),
and c) SNRs in the Galactic Disk. We discussed models a) and b) in \citet{cheng14} and \citet{cheng15}, respectively.

In \citet{cheng14}, we analysed a model of stochastic acceleration of electrons from the background plasma and showed that
the problem of the model is the effect of plasma overheating \citep[see][]{chern12}. However, for a specified set of acceleration
parameters the in-situ stochastic acceleration is able in principle to provide high energy electrons needed for the observed radio
and gamma-ray emissions from the FBs.

In \citet{cheng15} we analysed the hadronic model of gamma-ray emission from the FBs when gamma-rays are produced by $pp$
collisions while the radio flux is generated by secondary electrons.
Owing to low gas density in the halo, the efficiency of gamma ray production by $pp$ collisions in it is low.
\citep[In addition, a very long confinement of accelerated protons in the halo is needed $\sim 10^{10}$ years, see][]{crock11}.
We showed that in this model it is problematic to reproduce
the gamma-ray and radio fluxes from the FBs, and an additional component of primary electrons is necessary.
The magnetic field in this model is strongly restricted. The model that reproduces the observed gamma and microwave emission
from the FBs if the magnetic field is within the range $2.5$ to $7$ $\mu$G.

Below we analyse an alternative model of stochastic acceleration of electrons in the Bubbles. It is known that relativistic
electrons are produced by  supernova remnants (SNRs) which are distributed in the Galactic Disk. These electrons fill an extended region
(about several kpc above the plane) of the Galactic halo as found from radio and gamma-ray observations  \citep[see e.g.,][]{ber90,strong11}.
However, due to  inverse Compton and synchrotron energy losses, only  electrons with relatively low energies can penetrate into regions
high above the Galactic Plane while high energy electrons have shed most of their energy before they reach these regions.
The rate of energy losses is described as $dE/dt=-\mu E^2$ where the parameter $\mu$ depends on the density of background photons and
the strength of the interstellar magnetic field \citep[for details see e.g.,][]{ber90}.
For  the model of diffusion propagation of CRs presented, e.g., in \citet{acker}, the length scale of electron mean path length  for energies
$E=10^{12}$ eV is less than 1 kpc. Therefore, in order to produce the observed nonthermal emission from the FBs, these electrons
should be in-situ re-accelerated there.  An advantage of this model in comparison with the model of stochastic acceleration from
the background plasma with the temperature $\sim 2$ keV \citep[][]{cheng14} is that in the case the initial energy of accelerated electron
is already high, $\sim 1$ GeV, and the mechanism of re-acceleration needs to increase the energy of electrons penetrated into the Galactic halo
by three orders of magnitude only.

Charged particle in the FBs can be accelerated by scattering by MHD-waves \citep{mertsch} or by interaction with supersonic turbulence
\citep{cheng12}. These processes can be described as diffusion in the momentum space
\citep[for the formal equations of this process see e.g.,][]{topt85,byk92,byk93,ber90}.
We assume that the power necessary for the turbulence can be supplied by active processes in the Galactic Center (GC) when stars are captured by
the central supermassive blackhole. Energy as high as $W=10^{53}-10^{54}$ erg can be released by one capture \citep[see][]{cheng1,cheng2,cheng}.
We notice that even more energy can be released in the GC, $W\sim 10^{56}$ erg, if a giant molecular cloud is captured by the black
hole \citep[see][]{yang12,Zubo}. \citet{yus14} might have found tracers of the last capture of $10^5~M_\odot$ of gas which
occurred $\sim 10^{6.5}$ years ago.
Similar conclusion was obtained from UV data by  \citet{fox14}. They found indications on a strong outflow from the GC with the velocity
$\ga 900$ km s$^{-1}$ that might due to past activity of the GC over the last $\sim 2.5-4$ Myr. This time is comparable with the age of the FBs.

Below, we present our analysis of electron re-acceleration in the halo. In Section \ref{sec:plain_diff} we use the diffusion model to calculate
the flux of relativistic electrons emitted by SNRs in the Galactic Disk which reach altitudes of about several kpc and stochastic
re-acceleration of these electrons there up to energies about $10^{12}$ eV. In Section 3 we calculate the fluxes of radio and
gamma-ray emissions from the region of re-acceleration and compare these results with the data derived from the FB. In Section 4
we analyse the effect of convection transfer on spectra of accelerated particles and radiation. Section 5 provides a conclusion.

\section{The number and spectrum of re-accelerated electrons in the diffusion model}\label{sec:plain_diff}

In order to estimate the number of re-accelerated electrons and
their spectrum in the FBs, the kinetic equation should include
processes of particle propagation. The reason is that the electron
sources are in the Galactic Plane while acceleration processes
take place high above the Galactic Disk.  The kinetic equation for
the distribution function of electrons, $f(r,z,p)$, in this case is
\begin{eqnarray}
&&-\nabla \cdot\left[D(r,z,p)\nabla f -u(r,z)f\right]+\nonumber\\
&&\frac{1}{p^2}\frac{\partial}{\partial p}p^2\left[
\left(\frac{dp}{dt}-\frac{\nabla\cdot {\bf u}}{3}p\right)f -
\kappa(r,z,p)\frac{\partial f}{\partial p}\right] =
Q(p,r)\delta(z) \,,
\end{eqnarray} \label{eq_nu}
where $r$ is the galactocentric radius, $z$ is the altitude  above
the Galactic plane, $p=E/c$ is the momentum of electrons, $u$ is the
velocity of the Galactic wind, $D$ and $\kappa$ are the spatial
and momentum (stochastic acceleration) diffusion coefficients,
$c(dp/dt)=dE/dt=-\mu E^2$ describes the rate of electron energy losses, and $Q$
describes the spatial distribution of cosmic ray (CR)
sources in the Galactic plane ($z=0$) and their
injection spectrum. All parameters of this equation are discussed  in Appendix \ref{param}.

\begin{figure}[ht]
\begin{center}
\includegraphics[width=0.55\textwidth]{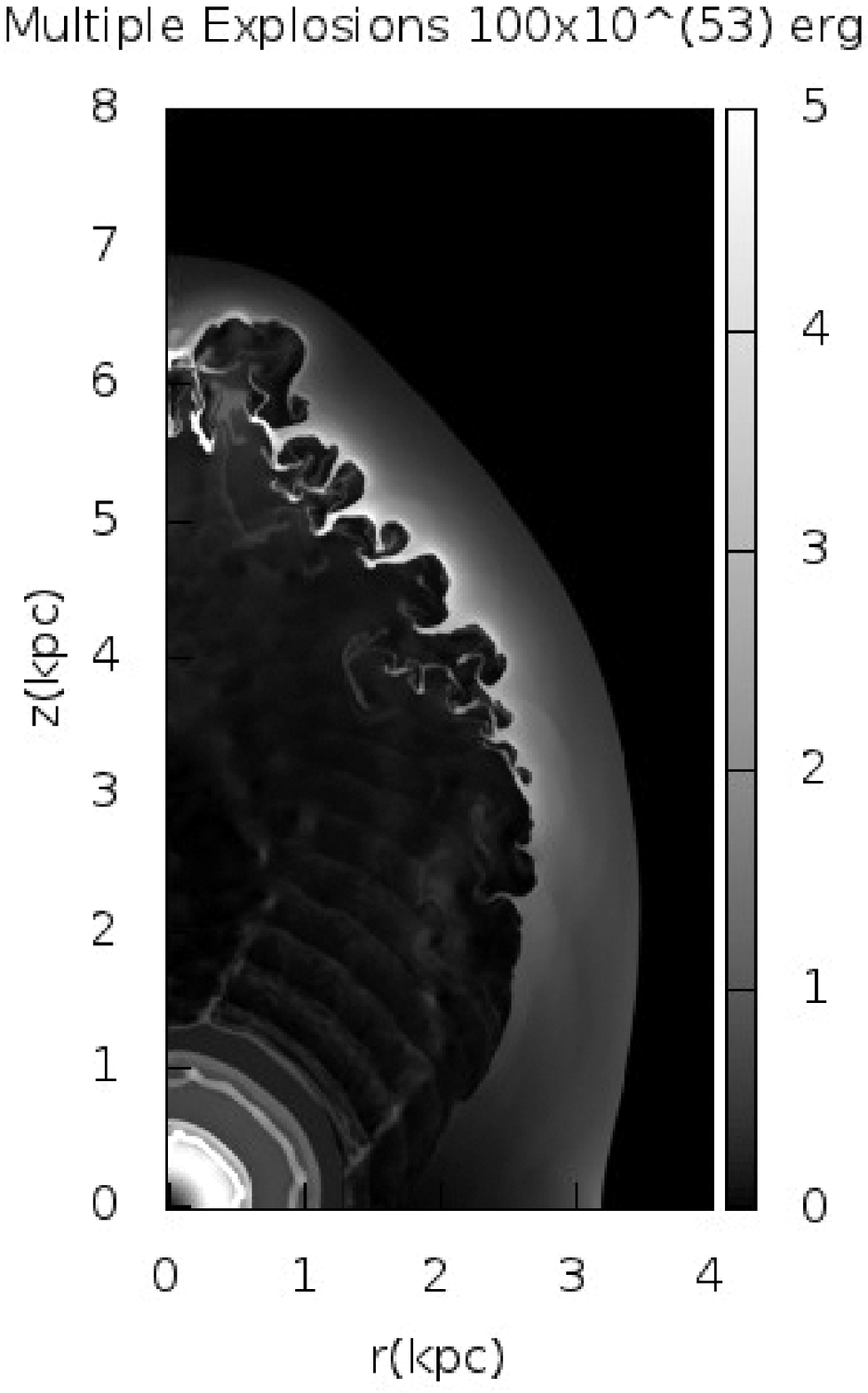}
\end{center}
\caption{Distribution of kinetic energy of gas from the hydrodynamic simulations of Fermi Bubbles.
In this simulation, we model the energy release by a star capture by SMBH at GC as an explosion at GC.
The figure shows the case of 100 captures with each energy release $10^{53}$ erg and
the interval between two successive captures is $10^5$ years.
The distribution is the result at $10^7$ years after the first capture.
Kinetic energy is plotted to emphasize the turbulent regions.
A layer of strong turbulence occurs close to the Bubble boundary.}
\label{struct}
\end{figure}

To define regions of stochastic acceleration or re-acceleration in the halo we used a hydrodynamic
code to simulate the propagation of energy released in the GC in an exponential atmosphere of the halo.
We adopted the code PLUTO \citep{Mignone07} and ran hydrodynamic simulations with cylindrical symmetry.
We described in the introduction that our idea on the formation of the FB is a series of star captures by
the SMBH at the GC \citep[e.g.,][]{cheng}.
We simulate the energy released by the capture as an explosion at the GC.
For illustrative purpose a typical result is shown in Fig. \ref{struct}.
Kinetic energy distribution of the gas is plotted in the figure to emphasize turbulent regions as we are going to discuss
stochastic acceleration processes in the bubbles.
This is the result of 100 captures with each energy release $10^{53}$ erg and the interval between
two successive captures is $10^5$ years.
The distribution shown is the results at $10^7$ years.

From the figure one can see that the shock propagation is mainly in the direction
perpendicular to the Galactic Plane and the morphology resembles the FB.
A layer of highly turbulent region is developed close to the envelope of the bubble.
The structure is similar to that excited by the Rayleigh-Taylor instability (RTI).
The development of the RTI at the shock front in an exponential atmosphere has been studied analytical by \citet{baumbr13}.

Similar effect of the RTI has been seen in SNR shocks \citep[see e.g.,][]{hester}.
Numerical calculations near a SNR shock provided by \citet{yang13} showed that the stochastic acceleration of electrons near a
SNR shock by magnetized turbulence may dominate over the shock acceleration because most energy of the magnetic fields
may be generated via the RTI. The total energy density of accelerated electrons in this case is of the order of energy
density of magnetic fluctuations. We cannot exclude that similar mechanism is effective behind the FB shock as \citet{mertsch}
assumed.

\begin{figure}[ht]
\begin{center}
\includegraphics[width=0.55\textwidth]{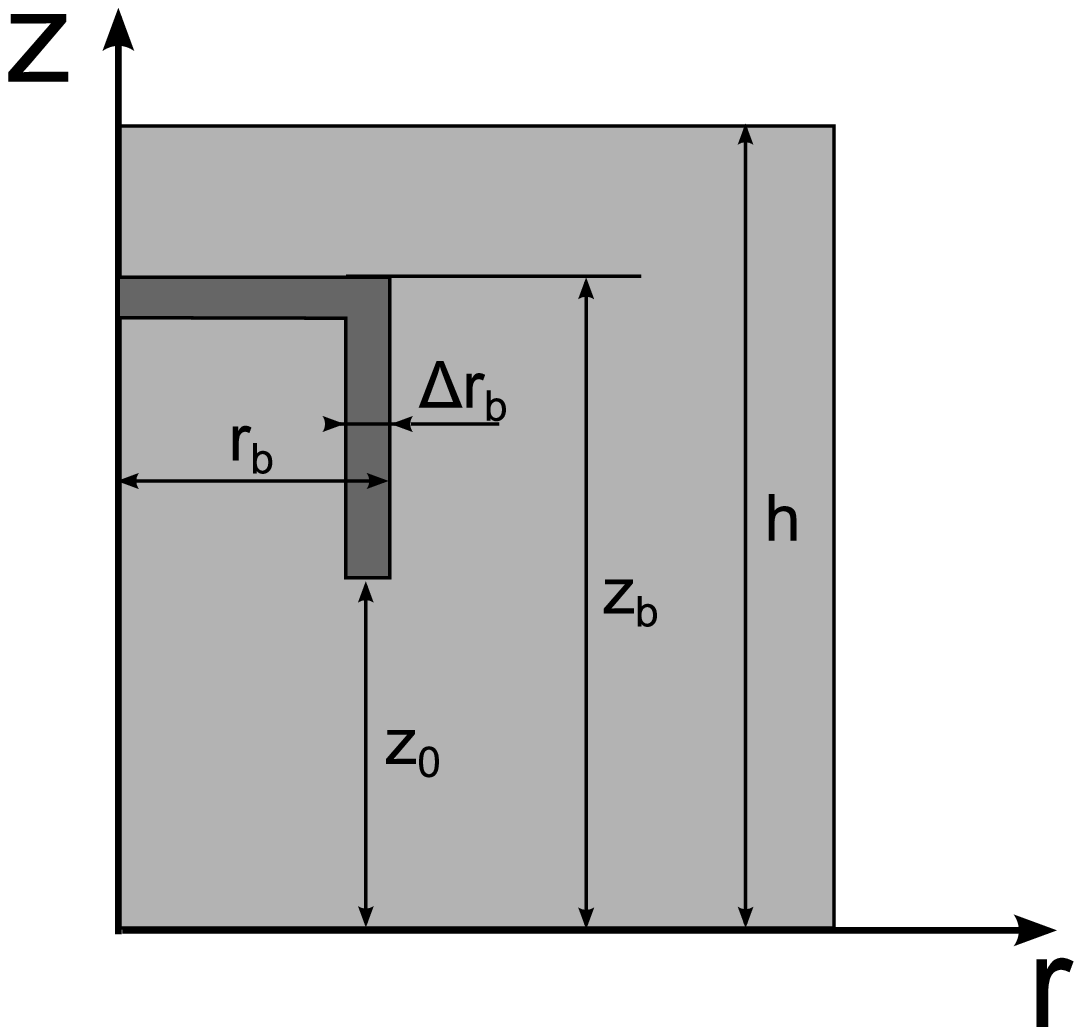}
\end{center}
\caption{The schematic picture of the Galactic halo (light gray) and the  re-acceleration
region of the FB (dark gray). In the model we put $h=8$ kpc, $r_b=3$ kpc, $z_b=5$ kpc and $z_0=3$ kpc.
The thickness of re-acceleration region, $\Delta r_b$ is estimated from the results of numerical calculations.}
\label{1056}
\end{figure}

To facilitate discussion and calculation,
we show schematically the region of electron re-acceleration in the FB where $\kappa\ne 0$ in Fig. \ref{1056}
by dark gray. For
calculations we used the model parameters taken from the GALPROP numerical program
\citep[][for details see Appendix \ref{param}]{acker}.
The momentum diffusion coefficient is taken in the form,
\begin{equation}
\kappa(p)=\alpha p^2\,.
\label{eq_dif}
\end{equation}

The parameter $\alpha$ can be presented as \citep[see e.g.,][]{ber90}
\begin{equation}
\alpha\sim \frac{c^2}{\nu}
\end{equation}
where $\nu$ is  the frequency of particle of scattering by magnetic fluctuations of the wave number $k$ is, e.g.,
\begin{equation}
\nu\simeq\omega_H\frac{\delta H(k)^2}{H_0^2}
\end{equation}
Here $\delta H(k)$ is the strength of magnetic  fluctuation with the wave number $k$, $H_0$ is the large scale
magnetic field, and $\omega_H=eH_0/m_ec$.

In the phenomenological model, which we investigate below, the
goal is to estimate the value of $\alpha$ from observational
data.

In the simplest case of pure diffusion  propagation of CRs  when the convection terms are
neglected ($u=0$), the number of electrons reaching the
re-acceleration region can be calculated from
\begin{equation}
-\nabla\cdot D(r,z,p)\nabla f + \frac{1}{p^2}\frac{\partial}{\partial
p}p^2\left[ \frac{dp}{dt}f - \kappa(r,z,p)\frac{\partial
f}{\partial p}\right] = Q(p,r)\delta(z) \,. \label{eq_nu_dif}
\end{equation}

The boundary conditions on the surface of the re-acceleration
region are rather questionable. The particle may freely escape
from the boundary as it is assumed for the Galactic halo
\citep[see e.g.][]{ber90}. On the contrary, \citet{yang12} assumed
the particle diffusion across the bubble edge is strongly suppressed, or there are ``magnetic walls''
at the edge as proposed by \citet{jones12}. Both effects prevent particle propagation through the bubble surface.

\begin{figure}[ht]
\begin{center}
\includegraphics[width=0.7\textwidth]{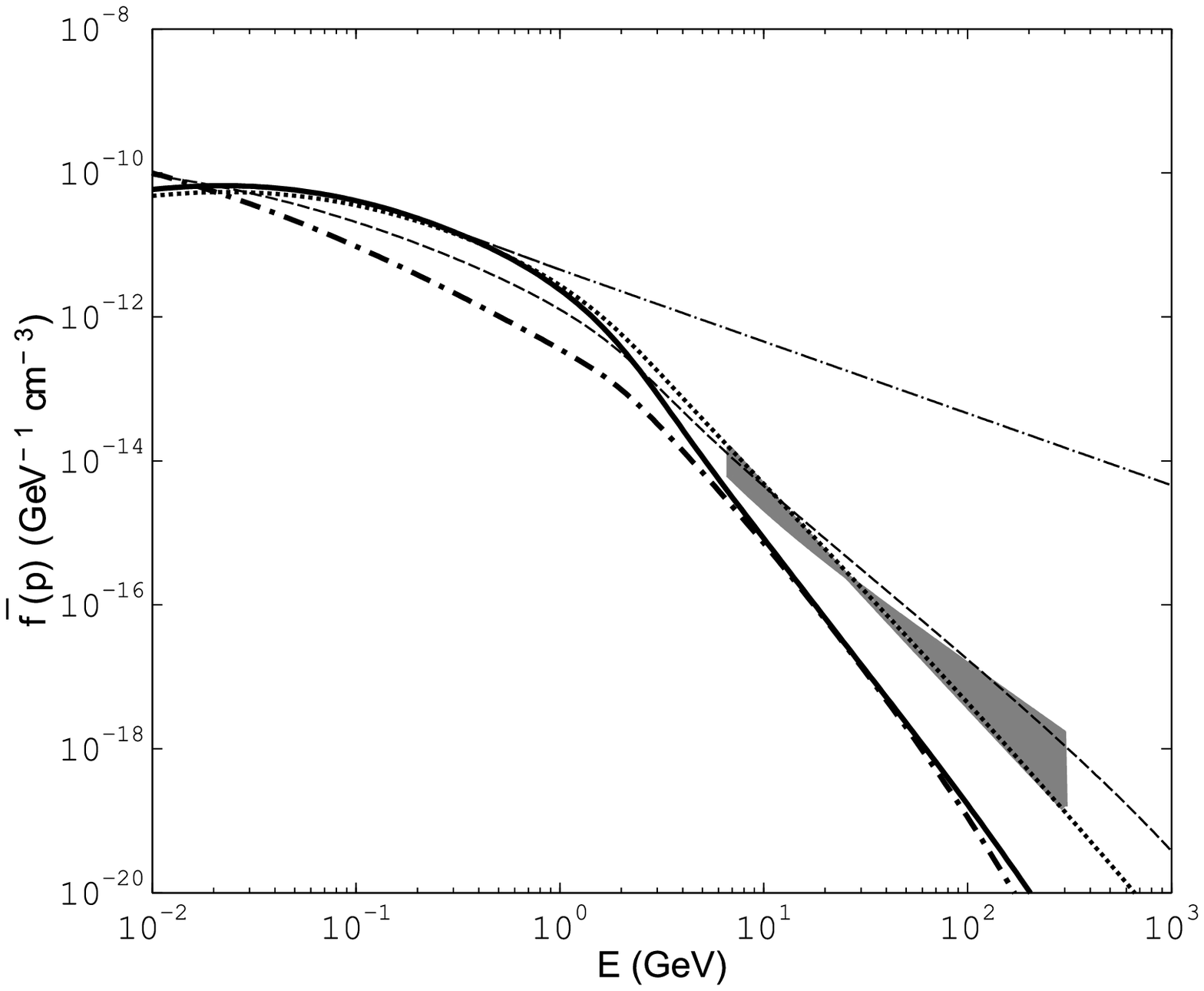}
\end{center}
\caption{The spectrum of SNR electrons accelerated in the
Bubbles at the altitude 5 kpc for the diffusion coefficient
$D=10^{29}$ cm$^2$ s$^{-1}$, $\mu=2\times
10^{-16}$ s$^{-1}$ GeV$^{-1}$ shown by the  thick solid line when convection term is neglected. The same spectrum of SNR
electrons penetrating into the re-acceleration region for the
convection model with the velocity gradient $v_0=10^{-15}$ s$^{-1}$ (as derived e.g. in \citet{bloe})
is shown by the thick dashed-dotted line. The re-accelerated
spectrum of electrons   in the Bubbles without escape term ($\propto E^{-1}$) is shown
schematically by the thin dashed-dotted line. The same  for the
case, when escape processes from the re-acceleration region of the thickness $\Delta r_b=3$ pc are
taken into account, is shown by the thin dotted line. The spectrum of re-accelerated
electron  when convection transfer is essential is shown by the thin dashed
line. The density of electron needed for the observed gamma-ray
flux from the Bubbles is shown by the gray region.}
\label{Fig_dif_conv}
\end{figure}

Below we accept for the boundary conditions of continuity for the particle density and flux.
From Eq.~(\ref{eq_nu_dif}) we calculated numerically the spectrum of SNR electrons which reach the altitude
$z=5$ kpc when the term of re-acceleration is neglected ($\kappa(p)= 0$). The spectrum  is shown in Fig.
\ref{Fig_dif_conv} by the solid line.

The density of high energy electrons needed for the observed gamma-ray flux is shown by the shaded region in Fig.
\ref{Fig_dif_conv}.
The necessary number of high energy electrons can be provided by processes of re-acceleration of SNR electrons in the FBs.
The spectrum of SNR electrons re-accelerated in the FBs is calculated from Eq. (\ref{eq_nu_dif}) when the acceleration term is
included ($\kappa(p)\neq 0$). However, with the acceleration in the form of Eq. (\ref{eq_dif}) the spectrum of accelerated particles
is too hard ($f(p)\propto p^{-3}$) that is shown schematically in Fig. \ref{Fig_dif_conv} by the dashed dotted line, i.e.,
our numerical calculations show that too many high energy electrons are produced in the re-acceleration region.

The spectrum of energetic particles can be steepened by processes of particle escape from the acceleration region
\citep[see e.g.,][]{cheng12}. Indeed, the momentum spectrum of accelerated particles is power-law,
$f(p)\propto p^{-\delta}$, with the spectral index $\delta$ given by
\begin{equation}
\delta = \frac{3}{2}
+\sqrt{\frac{9}{4}+\frac{\tau_{acc}}{\tau_{esc}}}\,,
\label{delta}
\end{equation}
where the acceleration time $\tau_{acc} \approx \alpha^{-1}$ and
escape time is $\tau_{esc} \approx \Delta r_b^2/4D_b$. Here
$\Delta r_b$ is the thickness of re-acceleration region and $D_b$
is the spatial diffusion coefficient in the re-acceleration region
given by
\begin{equation}
D_b(p) = {4v^2p^2}/(6\kappa(p))\,.
\label{dbp}
\end{equation}
Here $v$ is the velocity of turbulent motion. Escape processes make the spectrum steeper and thus decrease the number of
emitting electrons.

The spectrum of re-accelerated electrons in the Bubbles calculated for the model parameters derived from the observed FB
gamma-ray emission (see Section \ref{gamma_calc}) is shown in Fig. \ref{Fig_dif_conv} by the thin dotted line.
The ratio $\tau_{acc}/\tau_{esc}\simeq 8$ is shown in next section.

\section{Gamma-Ray and Radio Emission from the FB}\label{gamma_calc}

In order to calculate  the spectrum of accelerated particles in the Bubble, $f_b$, from Eq. (\ref{eq_nu_dif})
we calculated the total distribution function of electrons, $f(r,z,p)$, with the acceleration term ($\kappa\neq 0$).
Then we calculate the distribution function, $f_0(r,z,p)$  when the acceleration term is neglected ($\kappa= 0$).
We define the function the excess of electrons,  $f_b$, due to the acceleration as, $f_b=f-f_0$.
The procedure is similar to the subtraction of the FB gamma-ray flux from the  total Galactic emission \citep[see][]{meng,Yang2014}.

Then the FB gamma-ray intensity in the direction of the galactic coordinates $(\ell,b)$ is
\begin{equation}
I_\gamma(E_\gamma,\ell,b)=\frac{c}{4\pi}\int\limits_{s(\ell,b)}d
s\int\limits_\epsilon
n(\epsilon,r)d\epsilon\int\limits_{p}p^2f_b(r,p)\left(\frac{d^2\sigma}{d\epsilon~dp}\right)_{KN}dp\,.
\label{gamma}
\end{equation}
where $s(\ell,b)$ is the line of sight in the direction $(\ell,b)$, and $\left({d^2\sigma}/d\epsilon~dp\right)_{KN}$ is the
Klein-Nishina
cross-section \citep[see][]{blum}. We notice that the re-accelerated electrons fill not only the region of acceleration
but escape into the surrounding medium of the halo. Therefore, the total distance of emission is $l=\Delta r+\sqrt{D\tau}$
where $\tau$ is the lifetime of emitting electrons.

Our calculations show that in order to reproduce the FB gamma ray spectrum the following model parameters are required: the
spectral index of
accelerated electrons $\delta=4.8$, the thickness of the re-acceleration region $\Delta r_b=3$ pc, and the
characteristic time of acceleration $\alpha=2\times 10^{-13}$ s$^{-1}$.

\begin{figure}[ht]
\begin{center}
\includegraphics[width=0.8\textwidth]{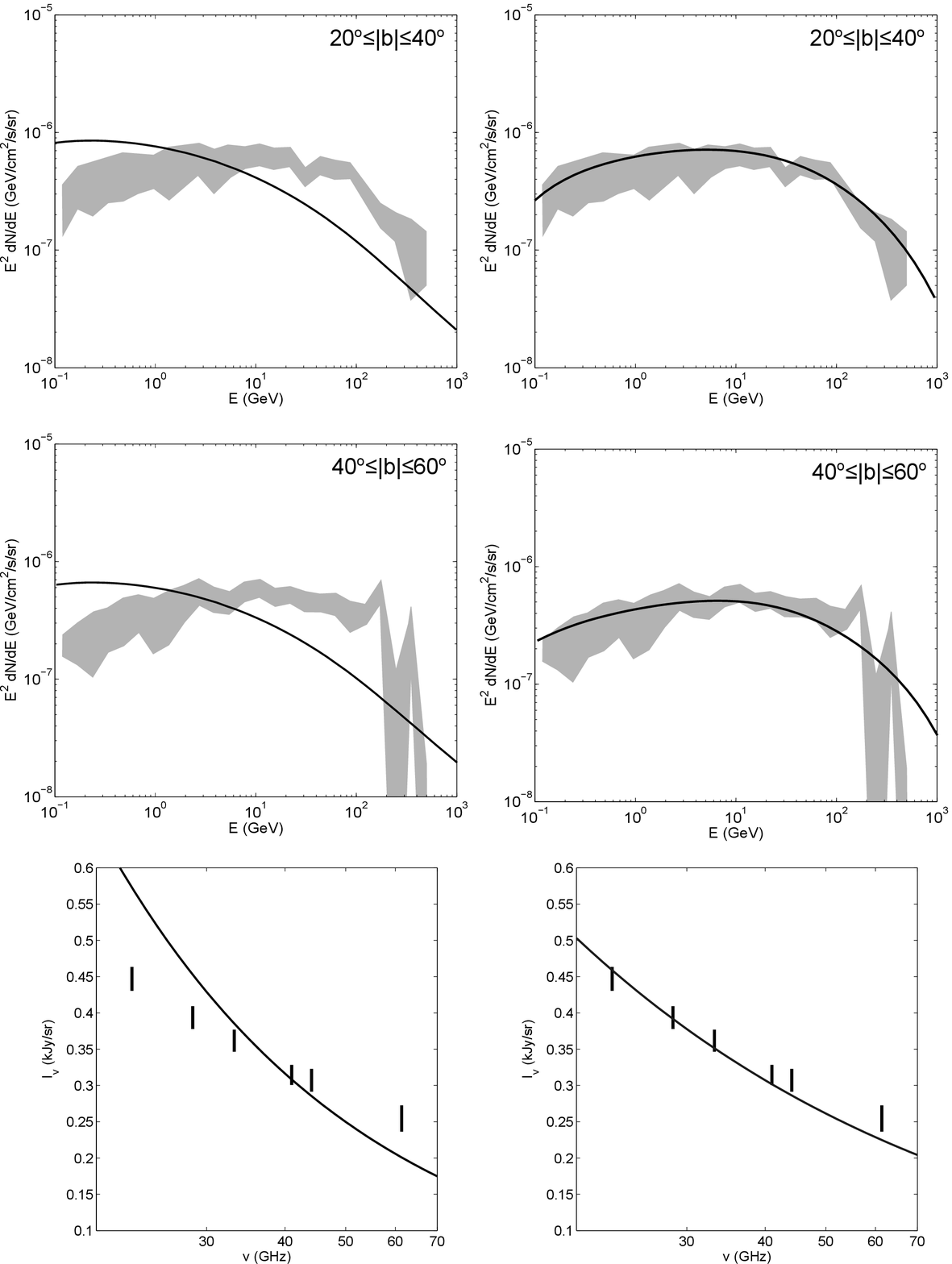}
\end{center}
\caption{Spectra  of gamma-ray and microwave emission produced by re-accelerated electrons in the FBs calculated for
the diffusion model (left column) and for the case when the effect of convection is included (right column).
The top and middle rows are gamma-ray emission for the latitude range $20^\circ\leq 40^\circ$ and $40^\circ\leq 60^\circ$, respectively.
The bottom row presents the spectrum of microwave emission from the Bubbles.
Datapoints were taken from \citet{acker14} for gamma and from \cite{ade13} for the microwave spectrum.} \label{spectr-lat}
\end{figure}

The calculated FB gamma-ray spectrum and the data from \citet{acker14} are shown in the top and middle rows of the left column panels
of Fig. \ref{spectr-lat}. As one can see there is no complete agreement between the calculation and the data.

This model has more serious problems to describe the microwave emission from the FBs.
The calculated  spectrum  is shown in Fig. \ref{spectr-lat} (bottom row, left column).
As one can see, for the  parameters derived from the gamma-ray data,
the model gives radio spectrum  steeper than $\nu^{-0.51}$  follows from measurements of Planck (see Fig. \citet{ade13}).

These shortcomings can be alleviated when CR convection propagation is included.
Below, we present a model of FB that takes convection into account.

\section{Effect of convection on the re-acceleration process}
The effect of convective transfer (Galactic wind) may also be essential in the Galaxy as it follows from theoretical treatments
\citep[see][]{breit91,breit} as well as from interpretation of observational data \citep[see][]{bloe}.

Analysis of the radio emission from the central Galactic region provided by \citet{crocker11} and
\citet{carr13} might indicate a very strong wind there. They estimated the wind velocity up to 1100 km s$^{-1}$.
Recent three-dimensional hydrodynamic simulations of \citet{mou14} showed the existence of strong winds in the FBs
caused by the past accretion in Sgr A*. They concluded that the wind is collimated by the Central Molecular Zone
towards the Galactic poles, i.e., perpendicular to the Galactic Plane.
Thus, we expect that the effect of wind transfer in the GC may not be negligible.

If the wind velocity is spatially nonuniform in the Galactic halo, then CRs
lose their energy by the adiabatic losses. Indeed, the MHD
numerical calculations of \citet{breit91} showed that the velocity
of Galactic wind increases almost linearly with the altitude $z$.
\citet{bloe} derived the value of the wind gradient from the cosmic
ray chemical composition. For the wind velocity in the form
$u(z)=3\upsilon_0 z$ they estimated the gradient value
$\upsilon_0$ as, $\upsilon_0\simeq 10^{-15}$ s$^{-1}$.

Formal solutions for the one-dimensional kinetic equation with the convection term are presented in Appendices B and C.

To demonstrate this effect we solved  Eq. (\ref{eq_nu}) with the both re-acceleration and convection terms.
According to \citet{crocker11} we assume that the wind is blowing mainly from the central part of the Galactic disk,
i.e., from the Bubble region. Therefore we took the wind velocity in the following simplified form
\begin{equation}
\label{uz_v0}
u_z = 3v_0z\theta(r_0-r)\,
\end{equation}
where $\theta(r)$ is the Heavyside function and $r_0 = 3$ kpc.

To define the spectrum of accelerated electrons we estimated the number of electrons that can reach the altitude about
several kpc (shown in Fig. \ref{Fig_dif_conv} by the thick dashed-dotted line).
Then we calculated the spectrum of electrons re-accelerated in the Bubbles.
The effect of adiabatic losses is shifting the spectrum as a whole into the range of smaller energies
(see the solution in Appendix \ref{conv_term} and an example in Fig. \ref{conv}) and that makes the spectrum of re-accelerated
particles flatter than in the diffusion model (see the thin dashed line in Fig. \ref{Fig_dif_conv}).

In the case of leptonic origin of the FB radio and gamma-ray
emission, which are  generated by  synchrotron and inverse
Compton energy losses of electrons respectively, the necessary
strength of magnetic field in the FBs can be estimated from the
simplified  equations for these processes presented in
\citet{ginz79}. For a power-law spectrum of relativistic
electrons, $N(E)=K_eE^{-\gamma_e}$ the magnetic field strength is
\begin{equation}
H\simeq\frac{1}{E_\gamma}\left[\frac{I_r}{I_\gamma}\cdot\frac{cw_{ph}\sigma_T}{2a(\gamma_e)}\cdot\frac{mc^2}{e^3}\cdot
\left(4/3\varepsilon\right)^{(\gamma_e-3)/2}\left(\frac{4\pi
mc\nu}{3e}\right)^{(\gamma_e-1)/2}\right]^{2/(\gamma_e+1)}
\label{h_est}
\end{equation}
which is independent from  the spectral parameter $K_e$ and
the  thickness of radiating region, if it  is the same for
the radio and gamma-ray emission. Here $a(\gamma_e)\simeq 0.1$,
$\sigma_T$ is the Thomson cross-section, $I_r$ and $I_\gamma$ are
intensities of radio and gamma-ray emission from the FBs,
$\varepsilon$ and $w_{ph}$ are  the energy and the energy density
of background photons in the FBs.

For the FB radio intensity $I_r=0.52$ kJy sr$^{-1}$ at the
frequency $\nu =23$ GHz and gamma-ray intensity $I_\gamma=4\times
10^{-9}$ ph cm$^{-2}$ s$^{-1}$ GeV$^{-1}$ sr$^{-1}$ at $E_\gamma=10$
GeV which is produced by scattering on optical or IR photons
\citep[see][]{cheng}  whose
energy density in the halo is about $w \simeq 0.2$ eV cm$^{-3}$,
we obtain for the electron spectral index $\gamma_e=2$ that the
magnetic field strength is
\begin{equation}
H\simeq 5~ \mu{\rm G} \label{h_min}
\end{equation}
This value is of the order of one derived below from more accurate
numerical calculations.

The procedure of calculating the
spectrum of accelerated particles $f_b$ is the same as described
in Section \ref{sec:plain_diff}.

The calculated spectra of gamma-ray emission at different
latitudes and radio from the Bubbles are shown in Fig. \ref{spectr-lat} (right column). The best
agreement with the data is achieved for $v_0 = 10^{-15}$ s$^{-1}$.
The magnetic field strength  is $H = 3\mu$G. The parameters of the acceleration are the following: the
thickness of the re-acceleration region is $\Delta r_b = 60$ pc
and $\alpha=2\times 10^{-14}$ s$^{-1}$.

We notice, however, that as it follows from Eq. (\ref{h_est}) it
is problematic to reproduce the gamma-ray spectrum in the
leptonic  model if we accept the magnetic field strength in the
Bubbles to be 15 $\mu$G as derived by \citet{carr13}. The density
of relativistic electrons estimated from the radio data is too low
in order to generate enough gamma-ray photons by inverse Compton
in the FBs.

\begin{figure}[ht]
\begin{center}
\includegraphics[width=0.8\textwidth]{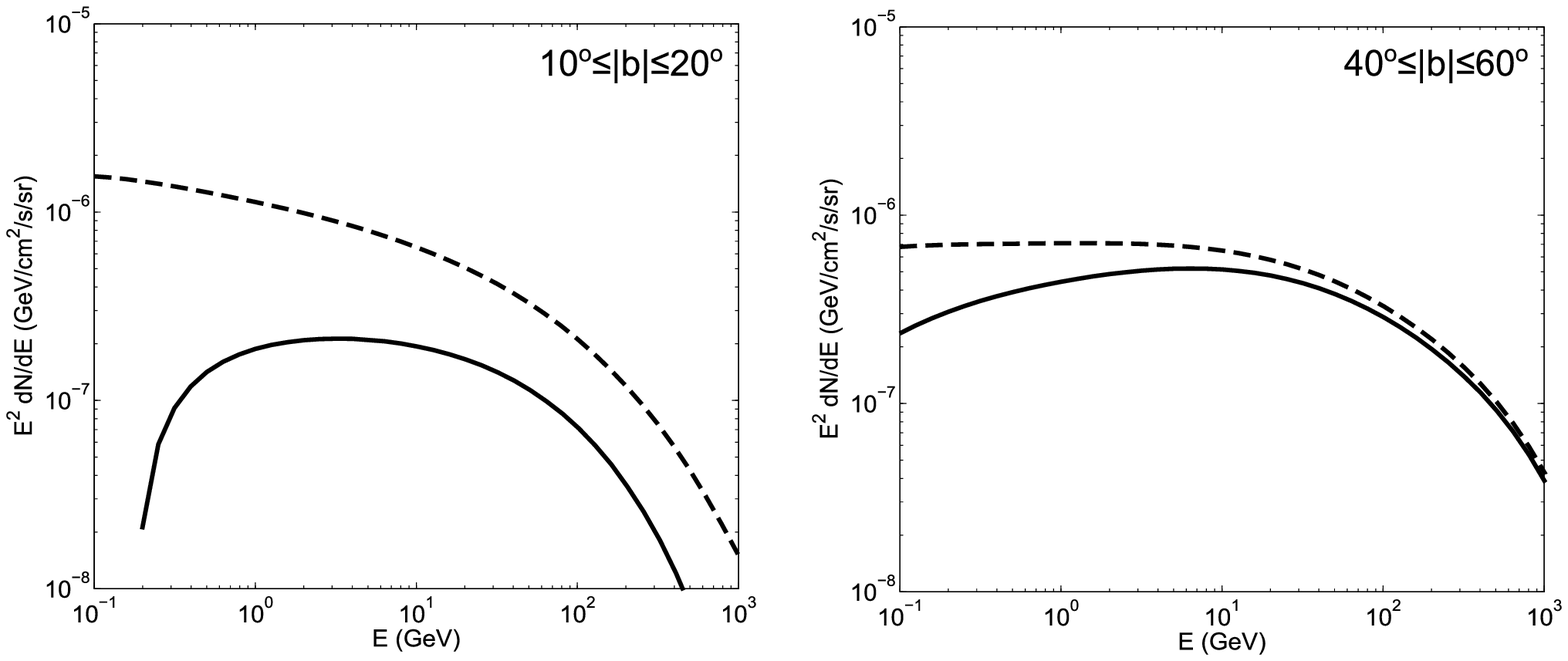}
\end{center}
\caption{Spectra of IC gamma-ray emission in the direction of low (left panel) and high (right panel) Galactic latitudes.
The spectrum of total IC emission (by FBs and SNRs of the Galactic Disk) is shown by the dashed line,
and the contribution by FBs only is shown by the solid line.}
\label{compar}
\end{figure}

At low Galactic latitudes the contribution of FBs should decrease especially in the high energy range
because of energy losses of electrons.
On the other hand, at these latitudes the contribution from electrons emitted by SNRs in the Disk increases.
This effect is shown in Fig. \ref{compar} where we present the gamma-ray spectrum in the direction of
low ($10^\circ<\mid b\mid<20^\circ$) and high ($40^\circ<\mid b\mid<60^\circ$) Galactic latitudes.
In this figure we show the IC component of gamma-ray emission produced only by the FBs (solid lines)
and the total IC emission produced by both FBs and SNR electrons (FB+SNR, dashed lines).
As one can see, in the framework of the model the contribution of the FBs to the total gamma-ray flux is significant
at high enough latitudes. At low latitudes the IC emission from SNR electrons is dominant and,
if we take also into account the gamma-ray component from proton-proton collisions, which is very intensive at low latitudes,
we conclude that it is almost impossible to subtract the FB component from the total gamma-ray flux in these directions.

The total power, $\dot{W}$ supplied by external sources (Fermi acceleration) which is needed to produce high energy electrons in
the FBs, can be estimated from \citep[see][]{chern12}
\begin{equation}
\dot{W}=-\int\limits_0^\infty\mathcal{E}\frac{\partial}{\partial p}\left(p^2\kappa(p)\frac{\partial f}{\partial p}\right)dp\,,
\end{equation}
where $p$ and $\mathcal{E}$ are the particle momentum and the kinetic energy, $f(p)$ is the particle distribution function,
and $\kappa$ is the diffusion coefficient of the Fermi acceleration.

We estimate  $\dot{W}$ from the observed FB gamma-ray and
microwave fluxes. We take into account all processes of
electron energy losses as well as their escape from the Galaxy (see Appendix \ref{param} for detail).
The power estimated numerically is about $\dot{W}\sim 2\times 10^{38}$ erg s$^{-1}$.

This estimate is a lower limit for $\dot{W}$, because a part of the energy released in the GC is also transformed
into accelerated protons and the plasma heating in the halo. As we showed in \citet{cheng} processes of star
accretion onto the central black hole can provide in average about $10^{41}-10^{42}$ erg s$^{-1}$. Thus tidal
accretion supplies enough energy for particle acceleration in the GC.

\begin{figure}[ht]
\centering
\includegraphics[width=0.55\textwidth]{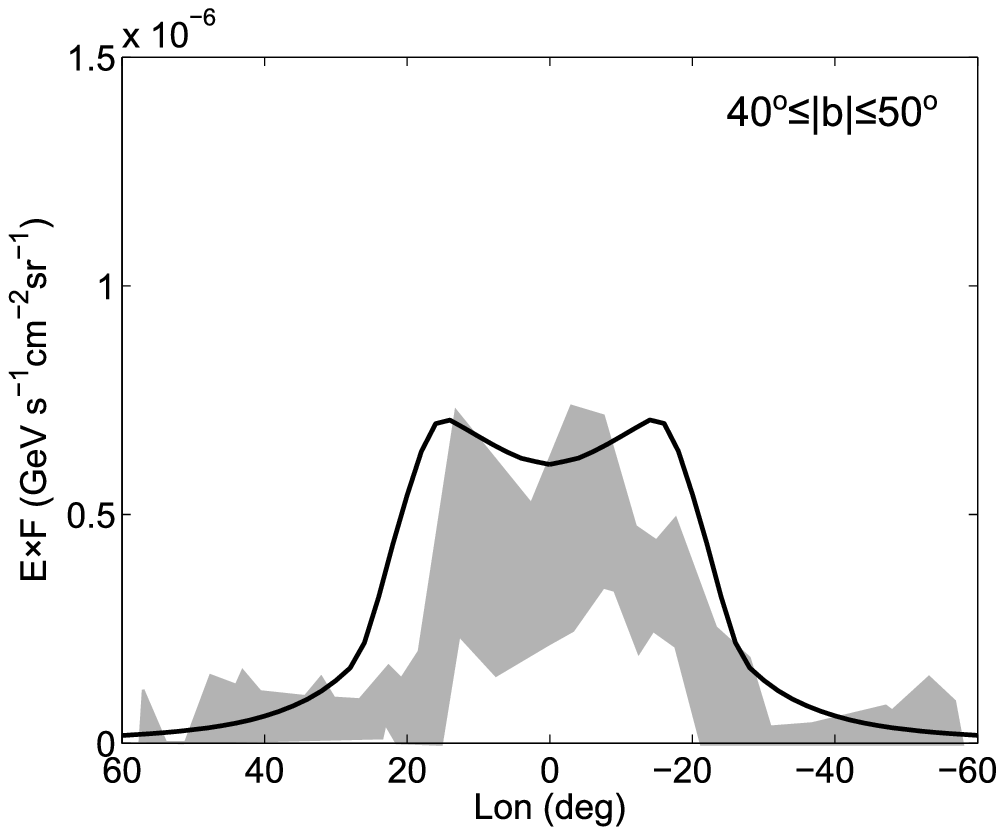}
\caption{Longitudinal distribution of the FB gamma-ray intensity for the latitude range
$40^\circ\leq b\leq 50^\circ$. Datapoints were taken from \citet{acker14}.}
\label{angular}
\end{figure}

Finally, we address the question whether the model is able to reproduce sharp edges of the Bubbles.
As an example we presented in Fig. \ref{angular} the longitudinal distribution of gamma-ray intensity.
As the geometry of the acceleration depicted in Fig. \ref{1056} is quite schematic,
we do not expect complete coincidence between the calculations and the data.
Nevertheless, the model reproduce qualitatively the effect of sharp edges and neither one-dimensional diffusion
nor magnetic walls at the bubbles edges are required \citep[in contrast to][]{yang12,jones12}.

\section{Conclusion}

In summary, we studied a leptonic model for the gamma-ray and microwave emissions from the FB.
The source of electrons is SNR in the Galactic Disk. The electrons propagate into the Galactic Halo and
are re-accelerated in the high turbulence regions in the FB which located several kpc about the Galactic Disk.
We have two goals to achieve. The first one is to get the density needed for the observed gamma-ray flux from the FB,
and the second one is to obtain the Planck microwave spectrum from the FB. We presented models of different level of sophistications.

To meet the first goal, the high energy electron density should be within the shaded area in Fig. \ref{Fig_dif_conv}.
We introduce the pure diffusion propagation model where we calculate the spectrum of electrons (from SNRs in the disk)
penetrating into the region of re-acceleration and calculate the spectrum of re-accelerated electrons.
The re-acceleration process generates very hard spectra ($E^{-1}$). Therefore electron density we obtained is too high
(the thin dashed dotted line above the shaded area in Fig. \ref{Fig_dif_conv}). We then introduce a fast particle escape
from the re-acceleration region. For small enough escape time the spectrum (thin dotted line in Fig. \ref{Fig_dif_conv})
passes through the shaded area in Fig. \ref{Fig_dif_conv}.

Although we got the density of electrons needed for the FB gamma-ray flux, the spectrum is too steep to produce the
Planck microwave spectrum from the FB ($\nu^{-0.51}$) and we fail the second goal. To remedy, we then include in the diffusion
with escape model the wind transport and adiabatic losses. The adiabatic losses shift the spectrum to lower energy range.
Adjusting the parameters of particle escape and wind in the halo we can obtain a flatter spectrum which satisfies both goals
(gamma-ray flux and Planck microwave spectrum).

\begin{figure}[ht]
\begin{center}
\includegraphics[width=0.6\textwidth]{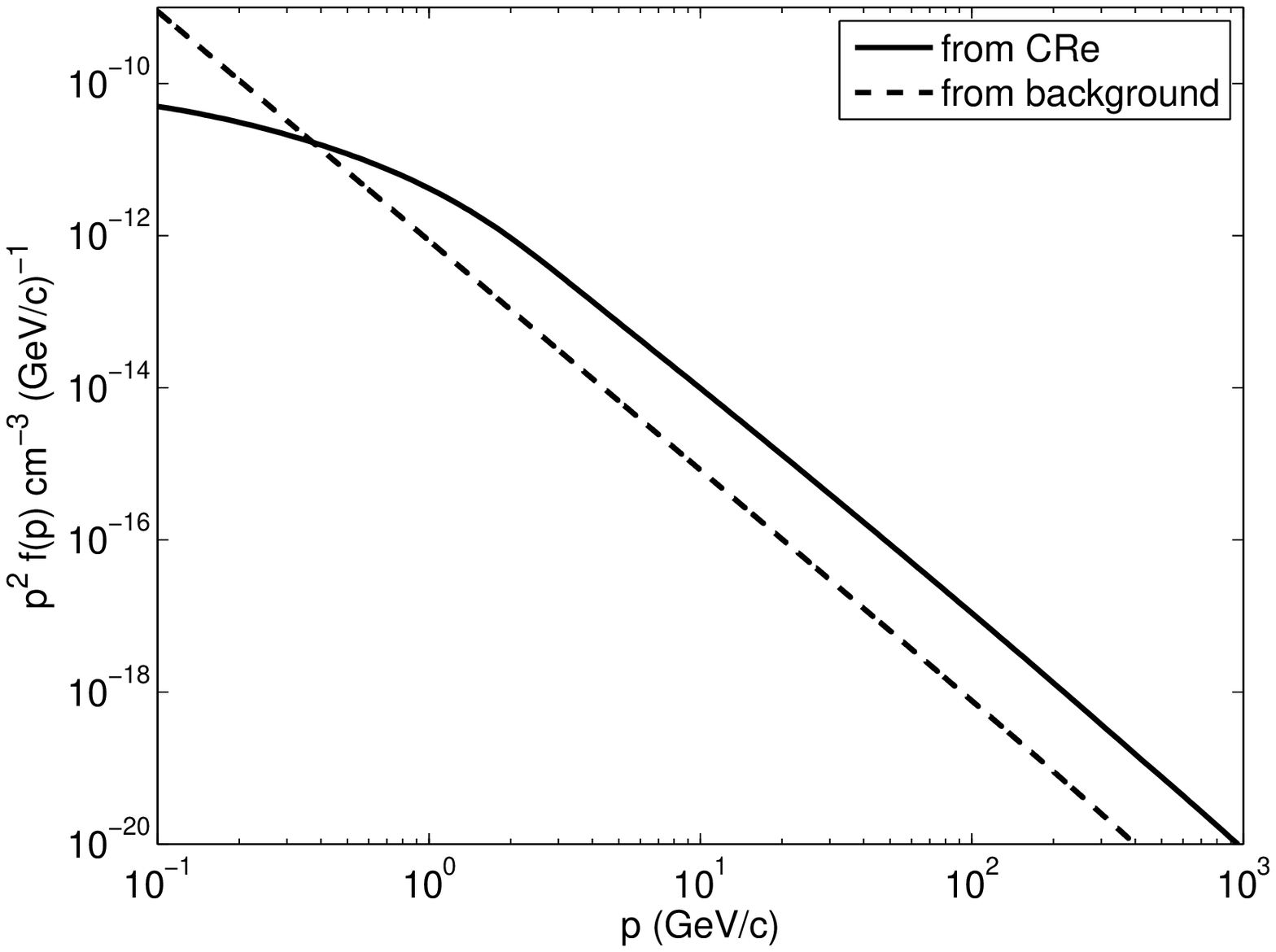}
\end{center}
\caption{Comparison of acceleration efficiency for the cases of
acceleration from a background plasma and for the case of
re-acceleration for the same parameters of momentum diffusion and
electron escape.} \label{acc_reac}
\end{figure}

The conclusions of the paper can be itemized as follows:

\begin{itemize}
\item{} Numerical calculations showed that the  energy of SNR electrons penetrating into the upper halo region is not high enough
to generate there the FB gamma-ray emission by the inverse Compton. Therefore, further re-acceleration up to energies about
$10^{12}$ eV is needed there to generate the gamma-ray flux.
\item{} Re-acceleration (without convection) generates too steep electron spectra. Therefore this model is unable to reproduce
correctly the gamma-ray and microwave emissions from the FBs.
\item{} There are indications on an intensive outflow of plasma from the GCÑ.
The effect of the wind leads to adiabatic losses of CRs. We expect that the adiabatic losses make the spectrum of electrons
in the acceleration region harder than for the case without convection. Our calculations show that the gamma-ray and radio
emissions of the re-accelerated electrons nicely reproduce the Fermi-LAT and Planck datapoints in this case.
\item{} In the re-acceleration model with convection, the gamma-ray flux produced by the FBs is more significant
at high Galactic latitudes than at low latitudes, see Fig. \ref{compar}. At low latitudes, the IC emission from
the FBs electrons is dominated by that from Galactic disk SNRs electrons.
Alternatively, gamma-rays can be produced by $pp$ collisions (hadronic model), but this process is not effective
in the Galactic halo where the gas and CR densities are low.
As a whole, at low latitudes the contribution of FBs to gamma-ray flux is subordinate to other processes.
We should point out that gamma-ray production by hadronic model (e.g., $pp$ collisions) could be distinguished from
leptonic model (e.g., inverse Compton process of electrons),
because $pp$ collisions also produce neutrinos \citep[see e.g.,][]{crock11,Lunardini2012,Taylor2014}.
\item{} An advantage of the re-acceleration model in comparison with that of acceleration from background plasma is that  in the first
case the energy of SNR electrons should be increased in the FBs by three orders of magnitude only, while in the second case electrons
are accelerated from the thermal plasma with the temperature about 2 keV, i.e., nine order of magnitude increase is needed.
\item{} We compared the efficiency of electron acceleration for the two models: acceleration from background plasma presented in
Paper I and the re-acceleration presented in this work. To do this we included both processes into the kinetic equation
(\ref{eq_nu_dif}) for electrons. Our numerical calculations showed that for the same momentum diffusion (acceleration term)
and the time of escape the re-acceleration mechanism is more effective for the production of high energy electrons in the FBs
(see Fig. \ref{acc_reac}).
\end{itemize}

\section*{Acknowledgements}

The authors are grateful to Y.W. Chang who performed the
hydrodynamic simulations of the Fermi Bubbles in Section \ref{sec:plain_diff}.
The authors thank the anonymous referee for valuable comments on an
earlier version of the paper.
KSC is supported by the GRF Grants of the Government of the Hong Kong SAR
under HKU 701013.
DOC is supported in part by the LPI Educational-Scientific Complex and
Dynasty Foundation.
DOC and VAD acknowledge support from the RFFI grants 15-52-52004, and 15-02-02358.
CMK is supported in part by the Taiwan Ministry of Science and Technology Grant
MOST  102-2112-M-008-019-MY3.
DOC, KSC, and VAD acknowledge support from the International Space Science
Institute to the International Team "New Approach to Active Processes in Central Regions of Galaxies".

\appendix

\section{Parameters of the CR kinetic equation}\label{param}

The distribution function of electrons, $f(r,z,p)$, is derived from the kinetic equation (see also Eq. (\ref{eq_nu}))
\begin{eqnarray}
&&-\nabla \cdot\left[D(r,z,p)\nabla f -u(r,z)f\right]+\nonumber\\
&&\frac{1}{p^2}\frac{\partial}{\partial p}p^2\left[
\left(\frac{dp}{dt}-\frac{\nabla\cdot {\bf u}}{3}p\right)f -
\kappa(r,z,p)\frac{\partial f}{\partial p}\right] =
Q(p,r)\delta(z) \,. \label{eq_nu1}
\end{eqnarray}

Relation between momentum and energy losses is
\begin{equation}
\frac{dp}{dt} = \frac{1}{\beta}\frac{dE}{dt} \,,
\end{equation}
where $\beta = v/c$ is dimensionless velocity of the electron and the rate of electron energy losses can be presented as
\citep[see e.g.,][]{ginz79}:
\begin{equation}
\frac{dE}{dt} = \left(\frac{dE}{dt}\right)_{cn} +
\left(\frac{dE}{dt}\right)_{ci} +
\left(\frac{dE}{dt}\right)_{br} +
\left(\frac{dE}{dt}\right)_{sc} \,,
\end{equation}
where the rate of Coulomb losses in a neutral medium and in a fully ionized plasma are respectively
\begin{equation}
\left(\frac{dE}{dt}\right)_{cn} = 7.62\times 10^{-18}\mbox{ GeV s$^{-1}$ } \times
n_{H}\beta^{-1}\left[\log(\gamma-1)(\gamma^2-1)+20.5\right] \,,
\end{equation}
\begin{equation}
\left(\frac{dE}{dt}\right)_{ci} = 7.62\times 10^{-18}\mbox{ GeV s$^{-1}$ } \times
n_{HII}\beta^{-1}\left[\log(\gamma/n_{HII})+73.6\right]
\end{equation}
Here $\gamma = {E}/(mc^2)=1/\sqrt{1-\beta^2}$ is the Lorenz-factor, $n_{H}$ and $n_{HII}$
are densities of neutral and ionized hydrogen in the disk and in the halo.

The bremsstrahlung losses  in the interstellar medium are
\begin{equation}
\left(\frac{dE}{dt}\right)_{br} = 5.1\times 10^{-19}\mbox{ GeV s$^{-1}$ } \times
(n_H + n_{HII})\gamma \,,
\end{equation}
and the  synchrotron and inverse-Compton losses (in the Compton limit) can be presented as
\begin{equation}
\left(\frac{dE}{dt}\right)_{sc} = 1.6\times 10^{-11}\mbox{ GeV s$^{-1}$ } \times
\left( H^2/8\pi +w_{sf}\right)\gamma^2\beta^2\,.
\end{equation}
The Klein-Nishina cross-section for inverse Compton scattering of high energy electrons can be found in \citet{blum}.

We take the values from the GALPROP code \citep[see][]{acker} for other parameters in Eq. (\ref{eq_nu1}).

The distribution of magnetic field in the Galaxy was taken in the from
\begin{equation}
H = 6~\mu\mbox{G}\times \exp\left[- \frac{z}{5\mbox{ kpc}} - \frac{r}{20\mbox{ kpc}}\right]\,,
\end{equation}
while the densities of neutral $n_{H}$ and ionized $n_{HII}$ as well as the
density of the interstellar radiation field $w_{sf}$ were taken from the GALPROP code of \citet{acker}.

The distribution of SNRs in the disk was taken in the form
\begin{equation}
Q(p,r) = Q(p) \times \left(\frac{r}{8\mbox{ kpc}}\right)^{1.2}
\exp\left(-3.22\frac{r}{8\mbox{ kpc}}\right)\,.
\end{equation}
As follows from the analyses of radio and gamma-ray the spectra of CR electrons and protons have a depletion at energies
below 1 GeV \citep[see e.g.,][]{strong11,neron12,dermer13}. Here we take the source spectrum of electrons as a
broken power-law from \citet{acker}
\begin{equation}
Q(p) \propto \left\{
\begin{array}{ll}
p^{-3.6}\,, & \mbox{ if } pc \leq 2.2 \mbox{ GeV}\\
p^{-4.4}\,, & \mbox{ if } 2.2\mbox{ GeV} < pc \leq 2.1\mbox{ TeV}\\
p^{-6}\,, &   \mbox{ if } pc > 2.1\mbox{ TeV}
\end{array}
\right. \,.
\end{equation}

The spatial  diffusion coefficients in the Galaxy is assumed to be a function of the momentum only,
\begin{equation}
D(p) = D_0 \times \beta\left(\frac{pc}{4\mbox{ GeV}}\right)^{0.33}\,,
\end{equation}
where  $D_0 = 9\times 10^{28}$ cm$^2$ s${-1}$ for 8-kpc halo.

Parameters inside the acceleration region were derived from the FB gamma-ray and radio emission.
The momentum diffusion coefficient has the form
\begin{equation}
\kappa_b(p) = {\alpha_b\over\beta}\, p^2\,,
\end{equation}
where acceleration rate $\alpha_b$ is estimated from calculations, and the spatial diffusion coefficient there is
\begin{equation}
D_b(p) = \frac{4v^2p^2}{6\kappa_b(p)}\,,
\end{equation}
where $v$ is the velocity of turbulent motions which provide stochastic acceleration.
In \citet{acker} this acceleration is due to particle interaction with Alfv\'enic waves, then $v=v_A=H/\sqrt{4\pi\rho}$.

The boundary conditions were taken in the form
\begin{equation}
\begin{array}{ll}
\partial f / \partial r = 0\,, & \mbox{ at }r = 0\\
\partial f / \partial z = 0\,, & \mbox{ at }z = 0\\
f = 0\,, & \mbox{ at the boundary of the Galactic halo.}
\end{array}
\end{equation}
At the boundary of the acceleration region, continuity of the particle function and flux are enforced.

\section{Re-acceleration in a divergent wind flow}\label{reacc-div}

In this appendix we present a solution to a one dimensional re-acceleration problem in divergent flow.
The governing equation is
\begin{equation}
  {\partial f\over\partial t}+u{\partial f\over\partial z}
  -{\!\partial\over\partial z}\left( D{\partial f\over\partial z}\right)
  -\left({\partial u\over\partial z}\right){p\over 3}{\partial f\over\partial p}
  -{1\over p^2}{\!\partial\over\partial p}
  \left(p^2\kappa{\partial f\over\partial p}\right)=Q\,.
  \label{re-acc-diverge}
\end{equation}
We consider a specific case where $u=u_0 z/H$, $D=D_0$, $\kappa=\kappa_0 p^2=\sigma^2p^2/9D_0$,
$Q=-f/\tau$, where $u_0$, $D_0$, $\kappa_0$, $\sigma$ and $\tau$ are constants.
$H$ is the characteristic length of the system or flow.
We seek steady state solution with boundary conditions:
$f=s(p)$ at $z=0$ and $f=0$ as $z\rightarrow\infty$.
Introducing dimensionless quantities
\begin{equation}
  \tilde{u}_0={Hu_0\over D_0}\,,\quad \tilde{\sigma}={H\sigma\over D_0}\,,\quad
  \tilde{\tau}={D_0\tau\over H^2}\,,\quad\xi={z\over H}\,,\quad\tilde{p}={p\over p_0}\,,\quad
  \eta=\int{3\over\tilde{\sigma}}{d\tilde{p}\over\tilde{p}}={3\over\tilde{\sigma}}\log \tilde{p}\,,
  \label{dimensionlessA}
\end{equation}
and
\begin{equation}
  f(\xi,\eta)=g(\xi,\eta)
  \exp\left[-\,{\eta\over 2\tilde{\sigma}}\left(\tilde{\sigma}^2+\tilde{u}_0\right)\right]
  =g(\xi,\eta)\exp(-\nu \eta)\,,
  \label{f-and-gA}
\end{equation}
the steady state of Eq.~(\ref{re-acc-diverge}) becomes
\begin{equation}
  {\partial^2 f\over\partial\xi^2}
  -\,\tilde{u}_0\xi{\partial f\over\partial\xi}
  +{\partial^2 f\over\partial\eta^2}
  +2\nu
  {\partial f\over\partial\eta}
  ={f\over\tilde{\tau}}\,,
  \label{steadystateEqA}
\end{equation}
and the boundary conditions for $g$ become:
$g=s(\tilde{p})\tilde{p}^{3\nu/\tilde{\sigma}}=S(\eta)e^{\nu\eta}$ at $\xi=0$ and $g=0$ as $\xi\rightarrow\infty$.
We can solve Eq.~(\ref{steadystateEqA}) by Fourier transform with respect to $\eta$.
We obtain an ODE
\begin{equation}
  {\partial^2 \bar{g}\over\partial\xi^2}-\,\tilde{u}_0\xi{\partial \bar{g}\over\partial\xi}
  -\left(\omega^2+\nu^2+{1\over\tilde{\tau}}\right)\bar{g}
  =0\,,
  \label{steadystateFourierA}
\end{equation}
subject to boundary conditions:
$\bar{g}=\bar{S}(\omega+i\,\nu)$ at $\xi=0$ and $\bar{g}=0$ as $\xi\rightarrow\infty$.
Here the Fourier transform pairs are
\begin{equation}
  g(\xi,\eta)=\int_{-\infty}^{\infty} \bar{g}(\xi,\omega)\,e^{i\,\omega\eta}\,d\omega\,,
  \quad
  \bar{g}(\xi,\omega)={1\over 2\pi}\int_{-\infty}^{\infty} g(\xi,\eta)\,e^{-i\,\omega\eta}\,d\eta\,.
  \label{Fourier-g}
\end{equation}
and
\begin{equation}
  S(\eta)=\int_{-\infty}^{\infty} \bar{S}(\omega)\,e^{i\,\omega\eta}\,d\omega\,,
  \quad
  \bar{S}(\mu)={1\over 2\pi}\int_{-\infty}^{\infty} S(\eta)\,e^{-i\,\mu\eta}\,d\eta\,.
  \label{Fourier-S}
\end{equation}
The solution of $\bar{g}(\xi,\omega)$ can be written in terms of parabolic cylinder function
\begin{equation}
  \bar{g}(\xi,\omega)={2^{-n/2}\over\sqrt{\pi\,}}\,\Gamma\left({1-n\over 2}\right)\,\bar{S}(\omega+i\,\nu)
  \exp\left({\tilde{u}_0\xi^2\over 4}\right)\,{\cal D}_{n}(\sqrt{\tilde{u}_0\,}\xi)\,,
  \label{Solution-gbar}
\end{equation}
where $\Gamma(x)$ is the gamma function, ${\cal D}_n(x)$ is the parabolic cylinder function, and
\begin{equation}
  n=-\,{1\over\tilde{u}_0}\left({\tilde{u}_0\over 2}+{\tilde{\sigma}^2\over 4}+{\tilde{u}_0^2\over 4\tilde{\sigma}^2}
  +{1\over\tilde{\tau}}+\omega^2\right)\,.
  \label{parameter-n}
\end{equation}
Define
\begin{equation}
  \bar{\cal{G}}(\xi,\omega)={2^{-n/2}\over\sqrt{\pi\,}}\,\Gamma\left({1-n\over 2}\right)
  \exp\left({\tilde{u}_0\xi^2\over 4}\right)\,{\cal D}_{n}(\sqrt{\tilde{u}_0\,}\xi)\,,
  \label{definition-G}
\end{equation}
and the Fourier transform pair
\begin{equation}
  {\cal{G}}(\xi,\eta)=\int_{-\infty}^{\infty} \bar{\cal{G}}(\xi,\omega)\,e^{i\,\mu\eta}\,d\omega\,,
  \quad
  \bar{\cal{G}}(\xi,\omega)={1\over 2\pi}\int_{-\infty}^{\infty} {\cal{G}}(\xi,\eta)\,e^{-i\,\omega\eta}\,d\eta\,.
  \label{Fourier-G}
\end{equation}
The solution can then be written as
\begin{equation}
  g(\xi,\eta)={1\over 2\pi}\int_{-\infty}^\infty S(\eta^\prime)\,e^{\nu\eta^\prime}\,{\cal{G}}(\xi,\eta-\eta^\prime)
  \,d\eta^\prime\,,
  \label{Solution-g}
\end{equation}
and
\begin{equation}
  f(\xi,\eta)={1\over 2\pi}\int_{-\infty}^\infty S(\eta^\prime)\,e^{-\nu(\eta-\eta^\prime)}\,{\cal{G}}(\xi,\eta-\eta^\prime)
  \,d\eta^\prime\,.
  \label{Solution-f}
\end{equation}

\section{ Analytical Solution of the One-Dimensional Wind Equation}\label{conv_term}

The one-dimensional equation for relativistic electrons can be presented in the form
\begin{equation}
-D_0E^\alpha{{\partial^2N}\over{\partial z^2}}+3v_0{\partial\over{\partial z}}\left(zN\right)
-{\partial\over{\partial E}}\left(\mu
E^2+v_0E\right)N=Q_0E^{-\gamma_0}\delta(z)\,.
\end{equation}
Here $D_0E^\alpha$ is the coefficient of electron diffusion in $z$ direction,
$V(z)=3v_0z$ is the wind velocity in $z$ direction,
$dE/dt=-\mu E^2$ is the rate of synchrotron and inverse Compton energy losses,
$Q_0E^{-\gamma_0}\delta(z)$ is the source function of electrons in the Galactic disk.

Introducing variables
\begin{equation}
\tau=\frac{\mu}{v_0 E}\,,
\quad
{\hat z}={z\over z_d}\,,
\quad
z_d=\sqrt{{D_0\over v_0}\left(v_0\over \mu\right)^\alpha\,}\,,
\quad
t=\int{{d\tau}\over{\tau^\alpha (1+\tau)^7}}\,,
\quad
\eta={{\hat z}\over{(1+\tau)^3}}\,,
\end{equation}
and the function
\begin{equation}
K={z_d v_0\over Q_0}\left({v_0\over \mu}\right)^{\gamma_0}{{(1+\tau)^4}\over{\tau^2}}\,N\,,
\end{equation}
we obtain the standard one dimensional diffusion equation
\begin{equation}\label{DiffusionEq}
{\partial\over{\partial t}}K-{{\partial^2}\over{\partial\eta^2}}K=F(t)\delta(\eta)\,,
\end{equation}
where $F(t)=\tau^{\gamma_0+\alpha-2}(1+\tau)^7$ with $\tau$ expressed as a function of $t$
by inverting $t=\int d\tau/\left[{\tau^\alpha (1+\tau)^7}\right]$.
The Green function of Eq.~(\ref{DiffusionEq}) can be found in \citet{morse53}.

\begin{figure}[ht]
\begin{center}
\includegraphics[width=0.8\textwidth]{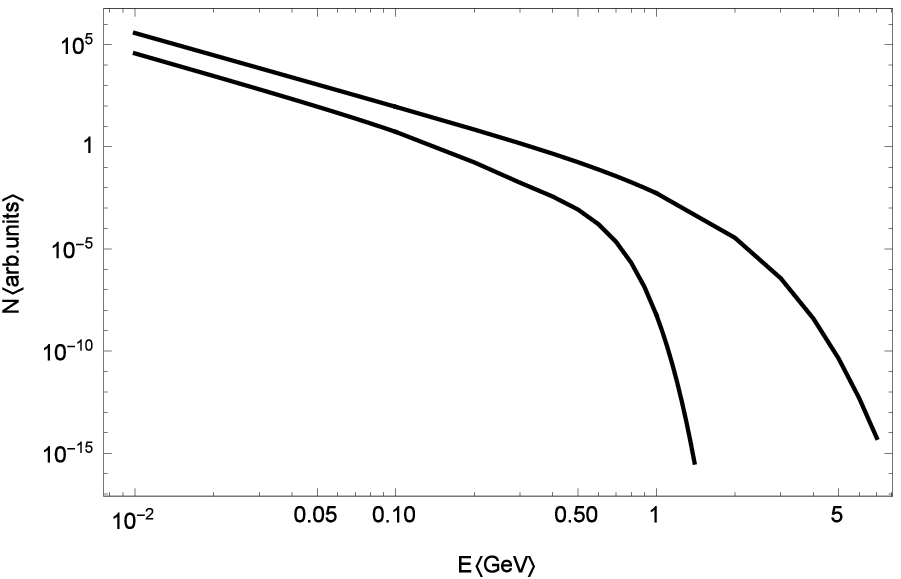}
\end{center}
\caption{This figure is an illustration of adiabatic loss in divergent flows.
Spectra of electron at the altitude $z=6.6$ kpc for the velocity gradients
$v_0=2\times 10^{-15}$ s$^{-1}$ (upper line) and $v_0=5\times 10^{-14}$ s$^{-1}$ (bottom line).}
\label{conv}
\end{figure}

The solution of this equation can be obtained in the analytical form
\begin{equation}\label{hat_n}
{N}={{Q_0h}\over{\sqrt{D_0v_0}}}
{{E^{-(\gamma_0+\alpha/2)}}\over{(1+ {{\mu E}/{v_0}})^4}}
{\int\limits_0^1}{d\tau_0\tau_0^{\gamma_0-2}\over \Sigma}
\exp\left[-\left({z\over \Sigma}\sqrt{{v_0}\over {D_0E^\alpha}}
{1\over{(1+{{\mu E}/{v_0}})^3}}\right)^2\right]\,,
\end{equation}
where
\begin{equation}
\Sigma={\int\limits_{\tau_0}^1}{{dx}\over{x^\alpha({{\mu E}/{v_0}}+x)^7}}\,.
\end{equation}
The effect of adiabatic losses is a shift of the spectrum into the region of low energies.
As an example we presented  in Fig. \ref{conv} the spectrum of electrons at the altitude $z=6.6$ kpc calculated
for the velocity gradients $v_0=2\times 10^{-15}$ s$^{-1}$ (upper line) and $v_0=5\times 10^{-14}$ s$^{-1}$ (bottom line).
The production spectrum of electrons by SNRs was taken from \citet{strong11} as
\begin{equation}
Q(E)=\left(\frac{4~{\rm GeV}}{E}\right)^{-1.6}\theta(4~{\rm GeV}-E)+\left(\frac{4~{\rm GeV}}{E}\right)^{-2.5}\theta(E-4~{\rm GeV}) \,.
\end{equation}

\begin{thebibliography}{99}
\bibitem[Ackermann et al.(2012)]{acker}
Ackermann, M., Ajello, M., Atwood, W. B.  et al., 2012, \apj, 750, 3
\bibitem[Ackermann et al.(2014)]{acker14}
Ackermann, M., Albert, A., Atwood, W. B. et al. 2014, ApJ, 793, 64
\bibitem[Ade et al.(2013)]{ade13}
Ade, P. et al.(Planck Collaboration)  2013, A\&A, 554, 139
\bibitem[Baumgartner \& Breitschwerdt(2013)]{baumbr13}
Baumgartner, V., \& Breitschwerdt, D. 2013, A\&A, 557, 140
\bibitem[Berezinskii et al.(1990)]{ber90}
Berezinskii, V. S., Bulanov, S. V., Dogiel, V. A., Ginzburg, V. L.,
\& Ptuskin, V. S. 1990, {\it Astrophysics of Cosmic Rays}, ed. V. L. Ginzburg, (Norht-Holland, Amsterdam)
\bibitem[Bland-Hawthorn \& Cohen(2003)]{cohen}
Bland-Hawthorn, J. \& Cohen, M. 2003, ApJ, 582, 246
\bibitem[Bloemen et al. (1993)]{bloe}
Bloemen, J. B. G. M., Dogiel, V. A., Dorman, V. L., \& Ptuskin, V. S. 1993, A\&A, 267, 372
\bibitem[Blumenthal \& Gould(1970)]{blum}
Blumenthal, G. R., \& Gould, R. J. 1970, Rev. Mod. Phys., 42, 237
\bibitem[Breitschwerdt et al. (1991)]{breit91}
Breitschwerdt, D., McKenzie, J. F., \& V\"{o}lk, H. J. 1991, A\&A, 245, 79
\bibitem[Breitschwerdt et al. (2002)]{breit}
Breitschwerdt, D., Dogiel, V. A., \& V\"{o}lk, H. J. 2002, A\&A, 385, 216
\bibitem[Bykov \& Fleishman(1992)]{byk92}
Bykov, A. M. \& Fleishman, G. D. 1992, MNRAS, 255, 269
\bibitem[Bykov \& Toptygin(1993)]{byk93}
Bykov, A. M. \& Toptygin, I. N. 1993, Physics Uspekhi, 36, 1020
\bibitem[Carretti et al.(2013)]{carr13}
Carretti, E., Crocker, R. M., Staveley-Smith, L. et al. 2013, Nature, 493, 66
\bibitem[Cheng et al.(2006)]{cheng1}
Cheng, K. S., Chernyshov, D. O. \& Dogiel, V. A. 2006, \apj, 645, 1138
\bibitem[Cheng et al.(2007)]{cheng2}
Cheng, K. S., Chernyshov, D. O. \& Dogiel, V. A. 2007, A\&A, 473, 351
\bibitem[Cheng et al.(2011)]{cheng}
Cheng, K.-S., Chernyshov, D. O., Dogiel, V. A., et al. 2011, ApJ, 731, L17
\bibitem[Cheng et al.(2012)]{cheng12}
Cheng, K  S., Chernyshov, D. O., Dogiel, V. A., et al. 2012, ApJ, 746, 116
\bibitem[Cheng et al.(2014, paper I)]{cheng14}
Cheng, K. S., Chernyshov, D. O., Dogiel, V. A., et al. 2014, ApJ, 790, 23, Paper I
\bibitem[Cheng et al.(2015, paper II)]{cheng15}
Cheng, K. S., Chernyshov, D. O., Dogiel, V. A., et al. 2015, ApJ, 799, 112, Paper II
\bibitem[Chernyshov et al.(2012)]{chern12}
Chernyshov, D. O., Dogiel, V. A. \& Ko, C.-M. 2012, ApJ, 759, 113
\bibitem[Crocker \& Aharonian(2011)]{crock11}
Crocker, R.M. \& Aharonian, F. 2011, PRL, 106, 101102
\bibitem[Crocker et al.(2014a)]{crocker14a}
Crocker, R. M., Bicknell, G. V., Carretti, E., Hill, A. S., \& Sutherland, R. S. 2014a, ApJ, 791, L20
\bibitem[Crocker et al.(2014b)]{crocker14b}
Crocker, R. M., Bicknell, G. V., Taylor, A. M. \& Carretti, E. 2014b, arXiv: 1412.7510
\bibitem[Crocker et al.(2011)]{crocker11}
Crocker, R. M., Jones, D. I., Aharonian, F., et al.  2011, MNRAS, 413, 763
\bibitem[Dermer et al.(2013)]{dermer13}
Dermer, C. D., Strong, A. W., Orlando, E., \& Tibaldo, L.; for the Fermi Collaboration 2013, arXiv1307.0497
\bibitem[Dobler \& Finkbeiner(2008)]{dob08}
Dobler, G., \& Finkbeiner, D. P. 2008, ApJ, 680, 1222
\bibitem[Dobler et al.(2010)]{dob10}
Dobler, G., Finkbeiner, D. P., Cholis, I., et al. 2010, ApJ, 717, 825
\bibitem[Finkbeiner(2004)]{fink}
Finkbeiner, D. P. 2004, ApJ, 614, 186
\bibitem[Fox et al.(2014)]{fox14}
Fox, A. J., Bordoloi, R., Savage, B. D., et al. 2015, ApJ, 799, L7
\bibitem[Ginzburg(1979)]{ginz79}
Ginzburg, V. L.  {\it Theoretical physics and astrophysics}, Oxford, Pergamon Press,
(International Series in Natural Philosophy. Volume 99), 1979
\bibitem[Hester et al.(1996)]{hester}
Hester, J. J., Stone, J. M., Scowen, P. A. et al. 1996, ApJ, 456, 225
\bibitem[Hooper \& Slatyer(2013)]{hoop}
Hooper, D., Slatyer, T.R., 2013, Physics of the Dark Universe, 2, 118
\bibitem[Jones et al.(2012)]{jones12}
Jones, D. I., Crocker, R. M., Reich, W., et al. 2012, ApJ, 747, L12
\bibitem[Lunardini \& Razzaque(2012)]{Lunardini2012}
Lunardini, C., \& Razzaque, S. 2012, PhRvL, 108, 221102
\bibitem[Mertsch \& Sarkar(2011)]{mertsch}
Mertsch P. \& Sarkar, S. 2011, PhRvL, 107, 1101
\bibitem[Mignone et al.(2007)]{Mignone07}
Mignone, A., Bodo, G., Massaglia, S., et al. 2007, \apj, 170, 228
\bibitem[Morse \& Feshbach(1953)]{morse53}
Morse, P. M., \& Feshbach, H. 1953, Methods of theoretical physics,
International Series in Pure and Applied Physics, New York: McGraw-Hill
\bibitem[Mou et al.(2014)]{mou14}
Mou, G., Yuan, F., Bu, D. et al. 2014, ApJ, 790, 109
\bibitem[Neronov et al.(2012)]{neron12}
Neronov, A., Semikoz, D. V., \& Taylor A. M. 2012, PhRvL, 108, 1105
\bibitem[Snowden et al.(1997)]{Snowden}
Snowden, S. L., et al. 1997, ApJ, 485, 125
\bibitem[Strong et al.(2011)]{strong11}
Strong, A. W., Orlando, E., \& Jaffe, T. R. 2011, A\&A, 534, 54
\bibitem[Su et al.(2010)]{meng}
Su, M., Slatyer, T. R., \& Finkbeiner, D. P. 2010, ApJ, 724, 1044
\bibitem[Taylor et al.(2014)]{Taylor2014}
Taylor, A. M., Gabici, S., Aharonian, F. 2014 PhRvD, 89, 103003
\bibitem[Toptygin(1985)]{topt85}
Toptygin, I. N. 1985, Cosmic rays in interplanetary magnetic fields, Dordrecht, D. Reidel Publishing Co.
\bibitem[Wardle \& Yusef-Zadeh(2014)]{yus14}
Wardle, M. \& Yusef-Zadeh, F. 2014, ApJ, 787, L14
\bibitem[Yang et al.(2012)]{yang12}
Yang, H.-Y. K., Ruszkowski, M., Ricker, P. M., et al. 2012, ApJ, 761, 185
\bibitem[Yang  \& Liu(2013)]{yang13}
Yang, C., \& Liu, S. 2013, ApJ, 773, 138
\bibitem[Yang et al.,(2014)]{Yang2014}
Yang R.-Z., Aharonian, F., Crocker, R., 2014, A\&A, 567A, 19
\bibitem[Zubovas \& Nayakshin(2012)]{Zubo}
Zubovas, K., \& Nayakshin, S. 2012, MNRAS, 424, 666

\end{thebibliography}
\end{document}